\documentstyle[psfig]{lamuphys}
\makeatletter
\let\chapter\hid@chapter
\makeatother
\begin{document}
\pagenumbering{arabic}
\pagestyle{empty}
\title{The Kormendy relation of HDF Ellipticals}
\author{Giovanni\,Fasano\inst{1}, Stefano\,Cristiani\inst{2} and
Stephane\,Arnouts\inst{2}}

\institute{Osservatorio Astronomico di Padova
\and
Dipartimento di Astronomia dell'Universit\`a di Padova}

\maketitle

\begin{abstract}

We present the results of the detailed surface photometry of a sample of
elliptical galaxies in the Hubble Deep Field.  In the $<\mu_e>$--$r_e$
plane the elliptical galaxies of the HDF turn out to follow a 
`{\it rest frame}' Kormendy relation, once the appropriate $K+E$
corrections are applied.  This evidence, linked to the dynamical information
gathered by \cite{st:et:al}, indicates that these galaxies,
even at $z \simeq 2-3$, lie in the Fundamental Plane, in a virial
equilibrium condition.  At the same redshifts a statistically
signifcant lack of large galaxies (i.e. with $\log r_e^{kpc} > 0.2$)
is observed.

\end{abstract}
 
\section{Data and techniques}

The present sample of {\it early--type} galaxies has been extracted
from the second release of the $WFPC2-HDF$ frames, in the $V_{606}$
band.  The selection is based on the photometry carried out by the
$ESO-STECF-HDF$ Group (\cite{cl:cou}) with the automated SExtractor
algorithm (\cite{be:ar}).

A preliminary list of candidate ellipticals was defined, including all
the galaxies satisfying the following criteria: {\it a)} Kron {\it
STMAG} magnitude in the $V_{606}$ band $\le$ 26.5; {\it b)} number of
pixels above the threshold limit of $1.3\sigma$ of the background
noise $\ge$ 200; {\it c)} star/galaxy classifier $(s/g)\ \le$ 0.6
($s/g$=1 means 'star').  The $\sim$400 objects matching the above
limits were examined to produce a first screening against late-type
objects.  The preliminary morphological classification was also
compared with the ones by \cite{vdb:et:al} and \cite{statler}, finding
a general good agreement.  Detailed surface photometry (luminosity and
geometrical profiles) was carried out on the resulting list of 162
{\it early--type} candidates.  As a consequence, 99 objects were left
in the final sample of 'bona-fide' {\it early--type} galaxies.
We derived the near infrared ($J$,$H$,$K$) total magnitudes of these
galaxies by applying the SExtractor algorithm to the deep images provided
by \cite{dick}.

We have restricted our analysis to the galaxies for
which a spectroscopic redshift is available (24), or a reliable
redshift estimate is possible (24), on the basis of multi--band
(optical {\it and} infrared) photometry.

Many galaxies in the present sample show a luminosity profile close to
the instrumental PSF. Therefore, in order to extract useful
morphological information, we used the '{\it Multi--Gaussian
Expansion}' deconvolution technique (\cite{bend}) to
restore the profiles.  Actually, \cite{fas:et:al} have shown that this
technique gives good results in recovering the '{\it true}' equivalent
half-light radius $r_e$ and the corresponding average surface
brightness $<\mu_e>$ down to $r_e^{true} \simeq FWHM$.

\section{The Kormendy relation of $HDF$ ellipticals}

The relation between the effective radius $r_e$ and the corresponding
average surface brightness $<\mu_e>$ (Kormendy 1985) is just a
projection of the Fundamental Plane of elliptical galaxies
(\cite{dj:da}).  The zero point of $<\mu_e>$ in the Kormendy relation
(hereafter $KR$) has been recently used in galaxy clusters at
different redshifts, as a standard candle to perform the classical
Tolman test for the expansion of the Universe and for testing the
evolutionary models of elliptical galaxies (\cite{pahre},
\cite{fas:et:al}). In the $HDF$ we are dealing with a
completely different situation, since the galaxies have very different 
redshifts. Still, putting the elliptical galaxies in the ($<\mu_e>$--$r_e$) 
plane is an interesting exercise.

\begin{figure}
\centerline{\psfig{figure=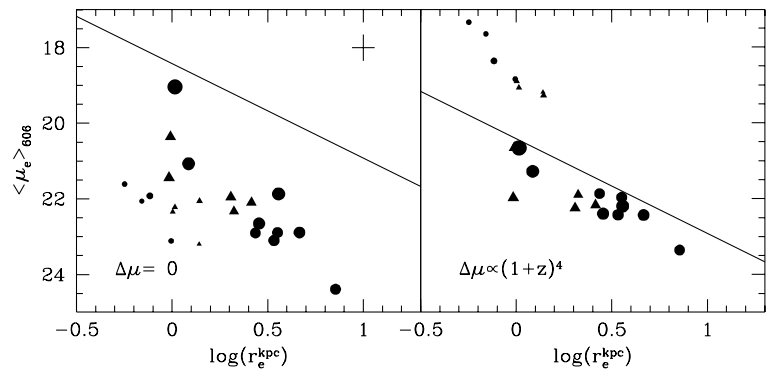,width=12.0cm,rheight=5.0cm}}
\caption[]{
The $<\mu_e>$--$r_e$ plane for HDF ellipticals with spectroscopic
redshift. In the left panel we plot the $<\mu_e>$ as derived from the
deconvolved profiles. The average error bars in both coordinates are
reported in the upper--right part of the plot. In the right panel we
have applied to $<\mu_e>$ the standard correction for the cosmological
dimming, $(1+z)^{4}$. The solid line represents the local $KR$ in the
$V_{606}$ band. The triangles represent the galaxies showing an inward
flattening of the luminosity profiles (with respect to the
de~Vaucouleurs' law) which cannot be ascribed to the effect of the
$PSF$ (see \cite{fil} and \cite{fas} for details).  The size of the 
symbols decreases with increasing redshift. 
}
\end{figure}

Figure 1 shows the $<\mu_e>$--$r_e$ plane for the 24 $HDF$ ellipticals
in our sample with spectroscopic redshift. By applying only the
$(1+z)^{4}$ cosmological dimming correction (right panel) we obtain
two distinct sequences for high redshift and low redshift galaxies in
the $<\mu_e>$ - $r_e$ plane. Both sequences are roughly parallel to
the local $KR$, with a small negative shift in the zero point at
low-redshift and a large positive shift at high redshift.  We also
note the tendency of high redshift galaxies ($z>2$) to be
intrinsically smaller with respect to galaxies with $z<1$.

In order to derive more realistic corrections for the observed surface
brightness of the $HDF$ ellipticals, we need to model their evolution.
We have used the chemical-spectrophotometric models of population
synthesis produced by \cite{bress} (`{\it closed-box}' models, $BCF$)
and \cite{tant} (models governed by the infall scheme, $TCBF$) to
estimate the K+evolutionary-corrections (K+E) for elliptical galaxies
of different mass, age and evolutionary history.

In order to verify which subset of the models is able to best describe
the elliptical galaxies observed in $HDF$ we have compared the
observed colors as a function of the (spectroscopic) redshifts with
the model predictions.  We found that a $BCF$ model with an `{\it
active}' star formation until the age of $11$ Gyr with a redshift of
formation $z_{form} = 5$ in a universe with $q_o = 0.5$ provides a
reasonable `global' description of the observed colors.  The {\it
passive} closed--box model and the models governed by the infall are
not able to reproduce the color--$z$ diagrams with a unique
redshift of formation. 

\begin{figure}[h]
\centerline{\psfig{figure=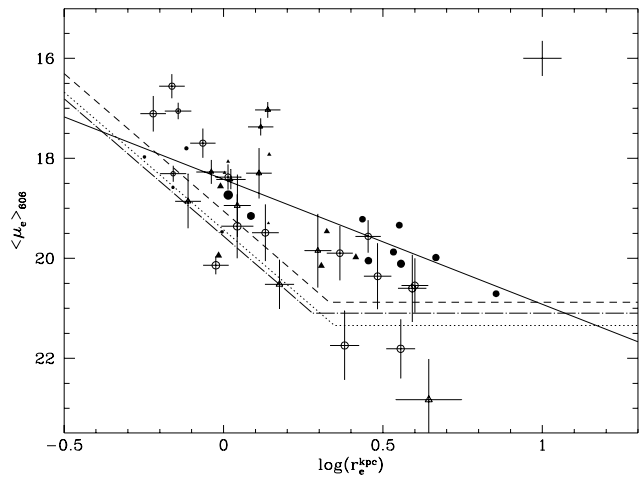,width=10cm,angle=-90}}
\caption[]{
The Kormendy relation for HDF ellipticals in our sample. The filled
and empty symbols refer to galaxies with spectroscopic and photometric
redshift, respectively. The symbols and the symbol sizes are the same
as in fig.1. The regions below the dotted, dashed and dot--dashed
lines represent, in the framework of the $BCF$ active models, the
forbidden regions of the Kormendy plane for z = 1, 2 and 3,
respectively (see text).
}
\end{figure}

To improve the statistics, we have estimated the photometric
redshifts of the 24 galaxies in the present sample having sufficient
and reliable infrared information (i.e. detection in $J$, $H$ and $K$,
no deblending problems), besides the optical magnitudes.

The Figure 2 indicates that, if we apply to the observed $<\mu_e>$ the
K+E-correction deduced from the galaxy models that properly reproduce
the observed colors ($BCF$ {\it active} model with $q_o=0.5$ and
$z_{form}=5$), the elliptical galaxies in the $HDF$ follow the local
Kormendy relation at any redshift.  In particular, for spectroscopic
redshifts (filled symbols), the agreement with the local $KR$ is
fairly good. On the other hand,
the $KR$ is well known to be a projection of the
Fundamental Plane which roughly represents the virial equilibrium
condition of galaxies. \cite{st:et:al} measured the line widths of
compact spheroidal galaxies at $z > 3$ which, if interpreted as due to
gravitational motions, imply velocity dispersions of the order $180
\le \sigma \le 320$ km s$^{-1}$.  Such values would place the high-z
galaxies of the present sample in the typical region of the
Fundamental Plane corresponding to ellipticals, showing that, even at
$z>2$, elliptical galaxies obey an average, rest--frame, virial
equilibrium condition.
On the other hand, the fact that the galaxies in our sample follow the
Kormendy relation is highly suggestive that a significant fraction of
the compact, high-surface brightness, high-z objects spectroscopically
observed by Steidel et al. (1996) have their lines widths due to
gravitational motions.

Fig.2 also shows that, high redshift galaxies are confined in the high
surface brightness, small effective radius region of the $KR$ plane.
Is this tendency due to the fact that large ellipticals have a faint
surface brightness (according to the Kormendy relation itself) ?
Large galaxies at high redshifts could lie below the detection limit.
We have checked at various redshifts what regions of the Kormendy
plane result ``forbidden'' because of the adopted selection criteria.
The regions of Fig.2 below the dotted, dashed and dot--dashed lines
represent, in the framework of the $BCF$ active models, the forbidden
regions of the Kormendy plane for z = 1, 2 and 3, respectively.  We
conclude that, had there been large galaxies at high redshifts in the
$HDF$, they would have been included in our sample.  In other words,
at $z>1.5$ a scarcity of large $r_e$ elliptical galaxies is observed
with respect to the local universe.

\end{document}